\documentclass[12pt]{article}
\usepackage{latexsym, amssymb, amsmath, theorem, epsfig, url}
\usepackage{natbib}
\usepackage{enumerate}
\usepackage{paralist}
\usepackage{bm}
\usepackage{float}
\floatstyle{ruled}
\newfloat{Algorithm}{t}{lox}
\floatname{Algorithm}{Algorithm}
\usepackage{algorithm}
\usepackage{algpseudocode}
\usepackage{booktabs}
\usepackage{subfig}

\newcommand{\blind}{0}

\newcommand{\qed}{ $\Box$ \medskip}
\newcommand{\Proof}{\noindent{\bf Proof.}~}
\newtheorem{theorem}{Theorem}
\newtheorem{proposition}{Proposition}
\newtheorem{lemma}{Lemma}

\renewcommand{\baselinestretch}{1.5}
\theoremstyle{break} 

\def\ba{{\boldsymbol{a}}}
\def\bk{{\boldsymbol{k}}}
\def\bv{{\boldsymbol{v}}}

\def\bmu{{\boldsymbol{\mu}}}

\def\btheta{{\boldsymbol{\theta}}}
\def\bsigma{{\boldsymbol{\sigma}}}

\date{}
\begin{document}

\def\spacingset#1{\renewcommand{\baselinestretch}%
{#1}\small\normalsize} \spacingset{1}


\if0\blind
{
\title{\bf Rating of players by Laplace approximation and dynamic modeling}
  \author{
    Hsuan-Fu Hua, Ching-Ju Chang, Tse-Ching Lin and
    Ruby Chiu-Hsing Weng\footnote{ \textit{chweng@nccu.edu.tw}}\hspace{.2cm}
   \\
    Department of Statistics\\
 National Chengchi University\\
 Taipei, Taiwan}
  \maketitle
} \fi

\if1\blind
{
  \bigskip
\begin{center}
\Large{\bf Rating of players by Laplace approximation and dynamic modeling}
\end{center}
  \medskip
} \fi

\bigskip
\begin{abstract}
The Elo rating system is a simple and widely used method for calculating
players' skills from paired comparisons data. 
Many have extended it in various ways.
Yet the question of updating players' variances remains to be further
explored.
In this paper, we address the issue of variance update by using the
Laplace approximation for posterior distribution,
together with a random walk model for the dynamics of players' strengths,
and a lower bound on players' variances.
The random walk model is motivated by the Glicko system, but here
we assume nonidentically distributed increments to take care of player 
heterogeneity.
Experiments on men's professional matches showed that the prediction 
accuracy slightly improves when the variance update is performed.
They also showed that new players' strengths may be better captured with the variance update.
\end{abstract}

\noindent%

\spacingset{1.45} 

\section{Introduction}
\label{sec:Intro}
The Elo system \citep{elo1978} 
was developed by Arpad Elo for rating chess players.
Let $\theta$ denote a player's strength and $\hat\theta$ be the estimated
strength. The system revises a player's skill by the formula:
\begin{align}
& \hat\theta' = \hat\theta + K(S-E),
\label{elo}
\end{align}
where $K$ is some fixed constant,
$S$ represents the actual game output, with $S=1$ if the player
wins and 0 otherwise, and $E$ is the expected outcome,
calculated as the expected win probability of the player. 
The update can be performed after each game or after a rating period.
The formula has an intuitive interpretation: 
the linear adjustment $K(S-E)$ is proportional to the difference
between the actual and the expected performance.
It is simple and very useful. 
Since its invention in the 1950s, the system and its 
modifications have been adopted by many organizations, including
the United States Chess Federation (USCF), the International
Chess Federation (FIDE), a variety of sports, and so on. 

Many have improved Elo by proposing model-based frameworks to 
incorporate the uncertainty in players' skills.
\cite{MEG93a,Glicko} and \cite{fahr94} 
proposed dynamic models that include time-varying parameters.
The Glicko system \citep{Glicko} is a Bayesian rating system that 
improves over Elo by incorporating the variability in parameter estimates.
There are also extensions of Elo to take additional domain information 
into account; see, for example,
\cite{hva10}, \cite{carb16} and \cite{koval2020} for the use of 
score differentials, or margins of victory.

A recent paper by \cite{ing21} 
showed that an approximate Bayesian posterior mode estimation 
gives an update rule resembling Elo’s formula.
He made the equal-variance assumption for players
and examined this assumption through experiments on predicting
men's tennis matches. 
Surprisingly, his experiments showed that
Glicko does not outperform standard Elo, despite its
ability to model changing uncertainty. 
He also considered a constant-variance version of Glicko and found that 
it performed slightly better than the standard Glicko.
That said, the equal-variance assumption is still unrealistic.
One reason is that as a player participates in more matches,
we would have more information about this player's rating, 
which in turn would be reflected in a lower variance.
The other reason is that the strength uncertainty could be heterogeneous
among players, and may even evolve over time; for instance, the uncertainty
about young competitors may be larger.
Though the Glicko system considers time-varying uncertainty,
it updates players' skills after a rating period
and \cite{Glicko} reports that the system performs best when the number of 
games per player is around 5-10 in a rating period.
Since updating after each match allows the ratings to adjust more quickly,
it is thus desirable to explore the estimation of players' variances after 
each match and integrate it into Elo's formula \eqref{elo}.
There are some related studies for variance update.
In the context of online games involving multiple teams and players,
TrueSkill \citep{RH06a,TTT08}, the ranking system for Microsoft Xbox Live, 
uses a factor graph and the approximate message passing to obtain the 
mean and variance update after each match. 
Yet it assumes that the strength follows the Gaussian distribution, 
corresponding to the Thurstone-Mosteller model \citep{Th27}, 
while the commonly used one is the Bradley-Terry model \citep{BR52a}.
\cite{wenglin} provide simple analytic update rules for mean and variance
based on moment matching, but 
their focus is also on multiple teams and players, and the evolution of
strength is not taken into account.

This paper addresses the variance update problem by using
the Laplace approximation for posterior distribution and employing
a random walk model for the strength evolution over time.
The Laplace approximation states that the posterior distribution 
can be approximated by a normal distribution whose mean is the posterior mode
and covariance is the inverse of negative Hessian of
the logarithm of posterior distribution.
With the Laplace approximation, the aforementioned Bayesian posterior 
mode approximation by \cite{ing21} can be extended to incorporate the
variance estimation.
The idea of describing players' strengths evolution by a random walk model
was motivated by Glicko, where the increments are assumed to be independent and 
follow a normal distribution with a mean of zero and a constant variance of $c^2$.
Therefore, between periods, the variance of a player's strength 
would increase by a constant $c^2$. 
Here we modify this approach by allowing the increments to be nonidentically
distributed. Explicitly, the variances of the increments are allowed to vary 
over time and across players. 
Random walks with nonidentically distributed increments have been studied
in the literature; see, for instance, \cite{denis18}.
We further impose a bound on players' variances to prevent it from
being overly low or high.
The use of a bound on the variance is not new; see, for instance,
the document for the Glicko system \citep{glickoDoc}.
We note that the Glicko-2 system \citep{glick01} also
assumes the strength evolves through a random walk model,
but the variance of the increment is assumed to be random
and time-varying, and the updating process requires iterative 
computation rather than involving only direct calculations like the 
Elo, the Glicko, and our proposed algorithms. 

Our variance update scheme can also be applied to
the GenElo Surface model \citep{ing21}. This model extended Elo 
by taking the effects of playing surfaces in tennis into account,
and it was shown to improve the prediction accuracy in professional 
men's tennis games. 
However, its update formula involves a matrix-vector product, 
which lacks intuitive interpretations. 
Specifically, the formula contains the term $H^{-1}J$, 
where $J$ and $H$ are the Jacobian
and Hessian of the logarithm of posterior distribution;
and in \cite{ing21}, $H$ is an $8 \times 8$ matrix whose inverse is not solved analytically.
We will solve $H^{-1}$ by a matrix inversion formula and rewrite Ingram's formula in an interpretive form.
Our formula reveals that the amount of adjustment to the strength of the unplayed 
surface is proportional to the correlation between the played and the 
unplayed surfaces, corrected by the ratio of standard deviations associated with
these two surfaces.
Furthermore, we remove the assumption of equal-variance for players
as in \cite{ing21} and show how the derived expression of $H^{-1}$ can be 
applied to update players' variances.

It is desirable to compare the prediction accuracy with and without using the proposed variance update scheme.
As the accuracy rates are calculated
from the same testing data, we conduct McNemar test for dependent proportions.
For an account of the test, we refer to \citet[Chapter 10]{agresti02};
see also \cite{diet98} for the use of McNemar test in comparing prediction 
performance of classifiers in machine learning areas.
For the experiments on men's professional matches,
we found that our variance update significantly improves the prediction accuracy. However, if the surface factor is incorporated into the model, the effect of variance update becomes insignificant, implying that the uncertainty of each player's
strength may be largely due to different playing surfaces. 
Therefore, we restrict our attention to the most played surface and the 
$p$-value is found to be around 0.05, indicating that the variance update 
is helpful.  
We also study the usefulness of the proposed variance update for new players.
It has been noted that the competitive ability for young players may
improve in sudden bursts. For example, \cite{simon97} proposed a model for
career trajectories and landmarks in the context of
creative productivity, and the aforementioned Glicko-2 system
addresses the fast improvement by a stochastic volatility model.
Our experiments showed that, for a majority of new players,
the use of variance update yielded higher prediction accuracy.

In summary, this paper aims to provide variance estimation for players' strengths
after each match. The variance formula is formed by three steps -- 
a basic formula based on negative inverse Hessian of the logarithm of the 
posterior distribution, 
an addition in variance to account for strength evolution over time, 
and a lower bound on players' variances. 
Our proposed method can be incorporated into Elo's rule
\eqref{elo} as well as the GenElo Surface model.
Moreover, we provide an intuitive interpretation for Ingram's formula.
Then, experiments on men's professional matches are conducted to assess 
prediction performance when variance update is performed.
In contrast to \cite{ing21}, which showed that the algorithm with
constant variance outperforms the ones with variance update,
we found that while the variance formula based on negative inverse Hessian performs poorly, higher prediction accuracy can be achieved
when time-varying strengths and a suitable bound on variance are employed. 
Our experiments also revealed that the variance update may better capture
new players' strengths in the early stage.
The organization of the paper is as follows.
Section 2 reviews rating systems relevant to the present paper.
Section 3 describes our proposed method.
Section 4 presents experiments on men's tennis data.
Section 5 concludes. Appendix contains some proofs and supplementary experiments.

\section{Review}
\subsection{Elo}
\label{subsec:Elo}
Denote the true ratings of players $i$ and $j$ as $\theta_i$ and $\theta_j$,
and their current estimates as $\hat\theta_i$ and $\hat\theta_j$.
Let $p_{ij}$ = P($i$ beats $j$), the win probability for player $i$, 
and let $s_{i}$ be the game outcome, where $s_{i}=1$ when player $i$ wins and 
$s_{i}=0$ when player $i$ loses. 
In Elo's system, each player's initial rating is set to 1500 and players' ratings
are updated by the formula:
$$\hat\theta^{'}_{i} = \hat\theta_{i} + K(s_{i}-\hat{p}_{ij}),$$
where $K$ is a value that has to be chosen in advance, and $\hat{p}_{ij}$ is 
the estimated win probability for player $i$. 
The currently implemented Elo system calculates $\hat{p}_{ij}$ by
$$
\hat{p}_{ij} = \frac{1}{1+10^{-(\hat\theta_{i} - \hat\theta_{j})/400}},$$
which corresponds to the model 
\begin{align}
p_{ij} = \frac{10^{\theta_{i}/400}}{10^{\theta_{i}/400} + 10^{\theta_{j}/400}} = \frac{e^{b\theta_{i}}}{e^{b\theta_{i}}+e^{b\theta_{j}}},
\quad \mbox{where } b = \log(10)/400.
\label{b}
\end{align}
The above equation is a reparametrized version of 
the Bradley-Terry model for paired comparisons \citep{BR52a}, which assumes that the win probability of $i$ over $j$ is $\lambda_i/(\lambda_i + \lambda_j)$,
where $\lambda_i>0$ is the strength of $i$.

We remark here that while
Elo's development of the chess rating system in late 1950s assumes
that a player's strength distribution follows a normal distribution,  
the model \eqref{b} can be derived by assuming that a player's strength
distribution follows an extreme value distribution. 
The paired comparison model derived from the normal distribution is known as 
the Thurstone-Mosteller model \citep{Th27}, 
but modern Elo and variants are mostly based on the Bradley-Terry paired comparison model.
See \citet[pp. 4-6]{glick95} for a detailed account.


\subsection{Glicko}
\label{gli}
The Glicko system \citep{Glicko} incorporated the variability in each 
player's rating. 
To the best of our knowledge, Glicko is the first Bayesian rating system.
It divides time into several rating periods,
where players' skills are assumed constant within each rating period.
Suppose that player $i$ has a strength parameter $\theta_{i}$ whose
prior distribution is $N(\mu_{i}, \sigma^{2}_{i})$. 
Assume the initial distribution for $\theta_i$ is $N(1500,\sigma_0^2)$, 
where $\sigma_0^2$ is the initial variance. 
Then, after observing game outcome data over a rating period, Glicko updates $(\mu_{i}, \sigma_{i}^{2})$ for each player based on large sample approximations.
Here the game outcome is modeled by the reparametrized Bradley-Terry model, given in \eqref{b}.
Let $s_{ij}$ be the game outcome between $i$ and $j$, 
with $s_{ij}=1$ if $i$ wins and $s_{ij}=0$ if $i$ loses;
let $E_{ij}$ denote the estimated win probability of $i$. 
Given player $i$'s current strength as 
$\theta_{it} \sim N(\mu_{it}, \sigma^{2}_{it})$
and opponent $j$'s strength as
$\theta_{jt} \sim N(\mu_{jt}, \sigma^{2}_{jt})$, $j=1,...,J$,
the Glicko algorithm is as follows:
\begin{align}
\begin{split}
&\mu_{i,t+1} =
\mu_{it} + \frac{b}{\frac{1}{\sigma^2_{it}} +\frac{1}{\delta_i^2}}
\sum_{j=1}^J g(\sigma_{jt}^2)(s_{ij}- E_{ij}),
\label{glicko}
\\
&\sigma^2_{i,t+1} =
\left(\frac{1}{\sigma^2_{it}} +\frac{1}{\delta_i^2}\right)^{-1} + c^2,
\end{split}
\end{align}
where $b=\log (10)/400$ as in \eqref{b},
\begin{align*}
& g(\sigma^2)=\frac{1}{\sqrt{1+\frac{3b^2 \sigma^2}{\pi^2}}},\\
& E_{ij}=\frac{1}{1+ 10^{-g(\sigma_{jt}^2)(\mu_{it}-\mu_{jt})/400}},\\
& \delta_i^2 = \left[b^2 \sum_{j=1}^J (g(\sigma_{jt}^2))^2 E_{ij} (1-E_{ji})\right]^{-1},
\end{align*}
and $c^2$ is a constant indicating the increase in the posterior variance 
from a rating period to the next rating period.
The addition of $c^2$ in \eqref{glicko} is justified by assuming that
strengths evolve over time through a random walk model,
\begin{align}
\theta_{it} = \theta_{i,t-1} + e_{it},
\label{rw}
\end{align}
where $e_{it} \sim N(0,c^2)$.
The calculations in \eqref{glicko} are carried out simultaneously for
each player over the rating period.
Both $c^2$ and the initial variance $\sigma_0^2$ are to be determined
in the model fitting process. 

The current and updated mean values of strength correspond to
the prior and posterior means of $\theta$. These values 
are usually denoted as $\mu$ and $\mu'$ \citep{Glicko} or
$\mu_{it}$ and $\mu_{i,t+1}$ \citep{glick18}.
The subscript $t$ is useful when a random walk like (\ref{rw}) is to be incorporated into the model.
In the rest of the paper, we will use $\mu'$ to denote the updated value.

\subsection{Ingram's approach}
\label{subsec:Genelo}
\indent \cite{ing21} presented an approximate 
Bayesian posterior mode estimation.
Let the ratings of the winner and loser be $\theta_w$ 
and $\theta_l$. Assume they have independent prior distributions $N(\mu_w, \sigma^{2})$ and $N(\mu_l, \sigma^{2})$. Let $\delta = \theta_w - \theta_l$ 
and $\tau = \theta_w + \theta_l$. 
Here the game outcome is also modeled by the reparametrized Bradley-Terry model in \eqref{b}.
So, the win probability of winner can be written as $\gamma(b \delta)$, where $\gamma(x) = (1+e^{-x})^{-1}$.
The logarithm of posterior distribution of $\delta$ given the game outcome is
$$\ell(\delta) = -\frac{(\delta - \mu_{\delta})^{2}}{2 \sigma_{\delta}^{2}} + \log \gamma(b\delta) + {\rm const},$$
where the posterior mean can be approximated by the posterior mode, and the mode can be approximated by a single Newton-Raphson step:
$$\mu'_{\delta} = \mu_{\delta} - \frac{\ell'(\mu_{\delta})}{\ell''(\mu_{\delta})}.$$
From this, the approximate posterior means 
$\mu_{w}'$ and $\mu_{l}'$ for the winner and loser are obtained.

For the incorporation of surface factor, suppose there are $n$ types 
of surfaces.
Let the ratings of players $i$ and $j$ be $\btheta_{i}$ and $\btheta_{j}$, 
and assume their prior ratings are
$\btheta_{i} \sim N(\bmu_{i},\Lambda)$ and 
$\btheta_{j} \sim N(\bmu_{j}, \Lambda)$,
where $\btheta_{i}$, $\btheta_{j}$, $\bmu_{i}$ and $\bmu_{j}$ are 
length $n$ vectors and $\Lambda$ is an $n \times n$ covariance matrix.
\cite{ing21} assumes $\Lambda$ to be the same across players and over time.
Suppose that $\btheta_{i}$ and $\btheta_{j}$ are independent. So,
their joint prior is 
$$
\btheta = \begin{bmatrix}
\btheta_{i}\\
\btheta_{j}
\end{bmatrix}
\sim N\left(
\begin{bmatrix}
\bmu_{i}\\
\bmu_{j}
\end{bmatrix},
\begin{bmatrix}
\Lambda& 0\\
0 & \Lambda
\end{bmatrix} 
\right)
= N \left(\bmu, \Sigma \right), \mbox{ say}.
$$
Define the vector $\ba$ as
$$\ba= 
\begin{bmatrix}
\ba_{i} \\
-\ba_{j}
\end{bmatrix},
$$
where $\ba_{i} = \ba_{j}$, an $n$-dimensional dummy vector with a 1 at the surface 
of the present game and zero otherwise. 
Now the win probability of $i$ is $\gamma( b \ba^{T}\btheta)$
and the logarithm of posterior distribution is
\begin{align}
\ell(\btheta) = -\frac{1}{2}(\btheta - \bmu)^{T} \Sigma^{-1}(\btheta - \bmu)+\log \gamma(b\ba^{T}\btheta) + {\rm const}.
\label{logp}
\end{align} 
Let $J(\btheta)$ and $H(\btheta)$ be the Jacobian and Hessian of 
$\ell(\btheta)$ in \eqref{logp}. Then, a single Newton-Raphson step is used
to update players' strengths:
\begin{align}
\bmu^{'} = \bmu - H^{-1}(\bmu)J(\bmu),
\label{bmu'}
\end{align}
where
\begin{align}
\begin{split}
& J(\bmu) = (1-\gamma(b \ba^T \bmu))b \ba,
\label{JHmu}
\\
& H(\bmu) = -\Sigma^{-1} - \gamma(b \ba^T \bmu)
(1-\gamma(b \ba^T \bmu)) b^2 \ba \ba^T.
\end{split}
\end{align}

Note that $H$ is a $2n \times 2n$ matrix and $J$ is a $2n$-dimensional vector
provided $n$ types of surfaces are considered.
In general, there are three primary types of tennis courts - grass, clay and hard,
but one may classify the surfaces into more than three types. 
For instance, \cite{ing21} considers four types - grass, clay, hard and indoor, which gives $n=4$ and $H$ an $8 \times 8$ matrix.

\section{Main results}
\label{main}
The Laplace method approximates the posterior distribution
of $\btheta$ with a normal distribution whose mean and covariance are the
posterior mode $\btheta^*$ and the inverse of negative Hessian matrix
$-H^{-1}(\btheta^*)$ of the logarithm of posterior distribution.
Let $\bmu$ be the prior mean of $\btheta$. It is natural to
extend Ingram's formula \eqref{bmu'} to include also
the covariance update; that is,
\begin{equation}
\begin{aligned}
\bmu' &= \bmu - H^{-1}(\bmu)J(\bmu),\\
\Sigma' &= -H^{-1}(\bmu').
\end{aligned}
\label{S'}
\end{equation}
Formula \eqref{bmu'} involves the $H^{-1}$, with the expression of $H$ in \eqref{JHmu}, yet the analytical forms of the elements of $H^{-1}$ are not solved in \cite{ing21}.
In Section \ref{revisit}, we will solve $H^{-1}$ by applying a matrix 
inversion formula and use it to write the matrix-vector product $H^{-1}J$ in
an interpretative form. 
With the expression of $H^{-1}$, we then present our variance update formula
of players' strengths in Section \ref{LapDyn}.
Note that $\Sigma'$ in \eqref{S'} is given by the negative inverse Hessian
evaluated at $\bmu'$ rather than $\bmu$, as the former should be closer to
the posterior mode.

\subsection{Revisited update rule for GenElo Surface model}
\label{revisit}
The GenElo Surface model \citep{ing21} extended Elo by taking the playing surfaces in tennis into account.
To re-express Ingram's update formula \eqref{bmu'} for the GenElo Surface model,
we shall allow the covariance of the strength $\btheta$ to vary across players
rather than assuming equal-variance as in \cite{ing21}.
Denote the covariance matrix of $\btheta_i$ as $\Lambda_{i}$.
So, the joint prior of $\btheta_{i}$ and $\btheta_{j}$ is
\begin{align}
\btheta = 
\begin{bmatrix}
\btheta_{i}\\
\btheta_{j}
\end{bmatrix}
\sim N\left(
\begin{bmatrix}
\bmu_{i}\\
\bmu_{j}
\end{bmatrix},
\begin{bmatrix}
\Lambda_{i} & 0\\
0 & \Lambda_{j}
\end{bmatrix} 
\right)
= N \left(\bmu, \Sigma \right).
\label{lamij}
\end{align}
If the game is played on surface $m$, then the win probability of $i$ is 
\begin{align}
p_{ij}(\btheta) = e^{b\theta_{im}}/(e^{b\theta_{im}}+e^{b\theta_{jm}}),
\label{im}
\end{align}
and the logarithm of posterior distribution in \eqref{logp} can be expressed as 
\begin{align}
\ell(\btheta) = -\frac{1}{2}(\btheta - \bmu)^{T}\Sigma^{-1}(\btheta - \bmu) +  s_{ij} \log (p_{ij}(\btheta)) + (1-s_{ij})\log (1-p_{ij}(\btheta)).
\label{t}
\end{align}
Straightforward calculation gives 
\begin{equation}
\begin{aligned}
J(\btheta) &= \nabla \ell(\btheta) = -\Sigma^{-1}(\btheta-\bmu) + 
\bv,\\
H(\btheta) &= \nabla^2 \ell(\btheta) = -\Sigma^{-1}+D.
\label{JH}
\end{aligned}
\end{equation}
Here $\bv=(v_1,...,v_{2n})^T$, with exactly two nonzero entries,
$v_m=b (s_{ij} p_{ji} - (1-s_{ij}) p_{ij})$ and
$v_{n+m}=-b (s_{ij} p_{ji} - (1-s_{ij}) p_{ij})$;
$D$ is a $2n \times 2n$ matrix whose $k$th row is the gradient
of $v_k$, $\nabla v_k$, and $D$ has four nonzero elements at the
$(m, m)$, $(m, n+m)$, $(n+m, m)$ and $(n+m, n+m)$ entries.
For instance, suppose that there are 
$n=3$ surfaces and $i$ beats $j$ on surface $m=3$. So, $s_{ij}=1$. 
Denote
\begin{align}
p_{ij}= p_{ij}(\btheta) \quad {\rm and}
\quad \hat{p}_{ij}=p_{ij}(\bmu),
\label{pim}
\end{align}
where $p_{ij}(\btheta)$ is in \eqref{im} and $\bmu$ is the prior mean of $\btheta$
as in \eqref{lamij}.
Then, we have $\bv = b(0, 0, p_{ji}, 0, 0, -p_{ji})^{T}$ and
\begin{align}
D = 
\left[
\begin{array}{c|c}
\phantom{1}D_1 & -D_1\\ \hline
-D_1 & \phantom{1}D_1
\end{array}\right]
\quad \mbox{with} \quad
D_1= 
\begin{bmatrix}
0	&\phantom{1}0	&\phantom{1}0\\
0	&\phantom{1}0	&\phantom{1}0\\
0	&\phantom{1}0	&-b^{2}p_{i}p_{j}
\end{bmatrix}
.
\label{D}
\end{align}
Note that the dependence of $\bv$ and $D$ on $\btheta$ is suppressed from the notation. 

In Proposition \ref{prop1} below,
the entries of the covariance matrices $\Lambda_i$ and $\Lambda_j$ in 
\eqref{lamij} are needed. Write 
\begin{align}
\Lambda_{i} = \begin{bmatrix}
\sigma_{11}	& \sigma_{12}	& \sigma_{13}\\
\sigma_{21}	& \sigma_{22}	& \sigma_{23}\\
\sigma_{31}	& \sigma_{32}	& \sigma_{33}
\end{bmatrix} 
\quad \mbox{and} \quad
\Lambda_{j} = \begin{bmatrix}
\tau_{11}	& \tau_{12}	& \tau_{13}\\
\tau_{21}	& \tau_{22}	& \tau_{23}\\
\tau_{31}	& \tau_{32}	& \tau_{33}
\end{bmatrix}.
\label{Lam}
\end{align}
Alternatively, by letting $\sigma_{ii}=\sigma_i^2$ and $\tau_{ii}=\tau_i^2$,
and denoting $\rho_{ij}$ as the correlation coefficient between surfaces
$i$ and $j$, we have
\begin{align*}
\sigma_{ij} = \rho_{ij}\sigma_i \sigma_j
\quad \mbox{and} \quad
\tau_{ij} = \rho_{ij}\tau_i \tau_j.
\end{align*}
In the following we may use the expressions 
$\{\sigma_{ij},\tau_{ij}: i,j=1,...,n\}$
or $\{\sigma_{i},\tau_{i}, \rho_{ij}: i,j=1,...,n\}$, whichever is more convenient.


\begin{proposition}
Suppose that there are $n$ types of surfaces and the current
game is played on surface $m$. 
Let $H$ be as in \eqref{JH} and $d=-b^{2}p_{ij}p_{ji}$. Then,
$-H^{-1}=\Sigma + W$, where
\footnotesize
\begin{align*}
&W= \frac{d}{{\rm det}(I-D\Sigma)}
\begin{bmatrix}
        \sigma_{1m}\sigma_{m1} & 
        \sigma_{1m}\sigma_{m2} & \cdots & \sigma_{1m}\sigma_{mn} & 
       -\sigma_{1m}\tau_{m1} & -\sigma_{1m}\tau_{m2} & \cdots & 
       -\sigma_{1m}\tau_{mn}\\
        \sigma_{2m}\sigma_{m1} &  \sigma_{2m}\sigma_{m2}  & \cdots &   \sigma_{2m}\sigma_{mn}  & 
      - \sigma_{2m}\tau_{m1}  & - \sigma_{2m}\tau_{m2}  & \cdots & 
      - \sigma_{2m}\tau_{mn} \\
      \vdots & \vdots & \ddots & \vdots & \vdots & \vdots & \ddots & \vdots\\
        \sigma_{nm}\sigma_{m1}  &  \sigma_{nm}\sigma_{m2}  & \cdots &  \sigma_{nm}\sigma_{mn} & -\sigma_{nm} \tau_{m1}  & 
        - \sigma_{nm}\tau_{m2}  & \cdots & - \sigma_{nm}\tau_{mn} \\
       -\tau_{1m}\sigma_{m1}  & - \tau_{1m}\sigma_{m2}  & \cdots &
       -\tau_{1m}\sigma_{mn}  &
         \tau_{1m}\tau_{m1}  &  \tau_{1m}\tau_{m2}  & \cdots &  \tau_{1m}\tau_{mn} \\ 
      -  \tau_{2m}\sigma_{m1}  & - \tau_{2m}\sigma_{m2}  & \cdots & - \tau_{2m}\sigma_{mn}  &
         \tau_{2m}\tau_{m1}  &  \tau_{2m}\tau_{m2}  & \cdots &  \tau_{2m}\tau_{mn} \\
         \vdots & \vdots & \ddots & \vdots & \vdots & \vdots & \ddots & \vdots\\
     - \tau_{nm}\sigma_{m1}  & - \tau_{nm}\sigma_{m2}  & \cdots & - \tau_{nm}\sigma_{mn}  &
        \tau_{nm}\tau_{m1}  &  \tau_{nm}\tau_{m2} & \cdots &   
        \tau_{nm}\tau_{mn} 
\end{bmatrix}
\end{align*}
\normalsize
and ${\rm det}(I-D\Sigma)=1-d(\sigma_{mm}+\tau_{mm}) 
=1+ b^2 p_{ij} p_{ji} (\sigma_m^2+\tau_m^2)$.
\label{prop1}
\end{proposition}
The result is derived by an application of a matrix inversion formula.
The proof is in the Appendix \ref{append:prop1}. 
Since $-H^{-1}$ is the approximate posterior covariance and $\Sigma$ is the
prior covariance, this proposition presents an explicit relationship
between the prior and posterior covariances.
The following proposition rewrites the matrix-vector product $-H^{-1}J$.
The proof follows from \eqref{JH} and Proposition \ref{prop1} and is omitted.
\begin{proposition}
Let $J$ and $H$ be as in \eqref{JH}. 
Suppose that there are $n$ types of surfaces and
the current game is played on surface $m$. Then,
$$
-H^{-1}(\bmu) J(\bmu)
= \begin{bmatrix}
\bk_{i}(s_{ij}-\hat{p}_{ij})\\
\bk_{j}(s_{ji}-\hat{p}_{ji})
\end{bmatrix},
$$ 
where $\hat{p}_{ij}$ is defined in \eqref{pim} and
$\bk_{i}$ and $\bk_{j}$ are $n$-dimensional vectors defined as
\begin{align*}
\bk_{i} 
=b C \sigma_m 
\begin{bmatrix}
\sigma_{1} \rho_{1m}\\
\sigma_{2} \rho_{2m}\\
\vdots\\
\sigma_{n} \rho_{nm}
\end{bmatrix}
\quad {\rm and} \quad
\bk_{j} 
=b C \tau_m
\begin{bmatrix}
\tau_{1} \rho_{1m}\\
\tau_{2} \rho_{2m}\\
\vdots\\
\tau_{n} \rho_{nm}
\end{bmatrix}
,
\end{align*}
with 
\begin{align}
C = (1+b^{2}\hat{p}_{ij}\hat{p}_{ji}(\sigma^{2}_{m}+ \tau^{2}_{m}))^{-1}.
\label{C}
\end{align}
\label{prop2}
\end{proposition}
Note that if there are $n$ types of playing surfaces, then
$\bk_i$ and $\bk_j$ in the above proposition will be $n$-dimensional.
The theorem below rewrites Ingram's formula \eqref{bmu'} in a form similar to 
Elo's update rule \eqref{elo}.
\begin{theorem}
Suppose that there are $n$ playing surfaces and 
that the current game is played on surface $m$. 
Let $\hat{p}_{ij}$ and $C$ be as in \eqref{pim} and \eqref{C}.
Then, the update of the mean skill for
player $i$ on surface $l \in \{1,...,n\}$ is
\begin{align}
\mu^{'}_{il} & = \mu_{il} + k_{il} (s_{ij}-\hat{p}_{ij}),
\label{muil}
\end{align}
where $k_{il} = bC \sigma_m \sigma_l \rho_{ml}$.
In particular, if $l=m$, we have $\rho_{mm}=1$ and $k_{im}=bC \sigma_m^2$. 
\label{thm1}
\end{theorem}
We make two remarks here.
First, if players' variances are assumed to be equal as in \cite{ing21},
then $\tau_l=\sigma_l$, for $l=1,...,n$, and \eqref{C} becomes
$C = (1+2b^{2}\hat{p}_{ij}\hat{p}_{ji}\sigma^{2}_{m})^{-1}$.
Second, the update rule \eqref{muil} resembles Elo's formula \eqref{elo} 
and enjoys a simple interpretation as follows.
Formula \eqref{muil} reveals that the amount of adjustment to the player's 
strength for the unplayed surface $l \neq m$ is
proportional to $\rho_{ml}$, the correlation coefficient between surfaces
$m$ and $l$. 
Further, if we write $\mu^{'}_{il} = \mu_{il} + \delta_{il}$ for $l=1,...,n$, 
where $\delta_{il}$ represents the adjustment to the strength for surface $l$,
then we have 
\begin{align}
\delta_{il}=\frac{\sigma_{l}}{\sigma_{m}}\rho_{ml} \delta_{im}.
\label{delta_il}
\end{align}
This means that the adjustment to the unplayed surface 
$l$ is proportional to the correlation between surfaces $m$ and $l$,
corrected by the ratio of standard deviations associated with
these two surfaces.
For instance, suppose that $\sigma_{grass}=100$, $\sigma_{hard}=80$,
and the correlation between grass and hard surfaces is 0.8;
after observing a game outcome on the grass surface,
if an update rule gives $\mu^{'}_{grass} = \mu_{grass} + 20$,
then, from \eqref{delta_il}, the update for the hard surface is
$\mu^{'}_{hard} = \mu_{hard} + (80/100)*0.8*(20) = \mu_{hard}+12.8$.

\subsection{Update via Laplace approximation and dynamic modeling}
\label{LapDyn}
Recall that the Laplace approximation suggests the update formula \eqref{S'}:
\begin{align*}
\bmu' &= \bmu - H^{-1}(\bmu)J(\bmu),\\
\Sigma' &= -H^{-1}(\bmu').
\end{align*}
The expressions of $-H^{-1}$ and $-H^{-1}J$ for 
the GenElo Surface model \citep{ing21} are given in 
Propositions \ref{prop1} and \ref{prop2}.
In particular, if the surface factor is not taken into account, meaning
that all surfaces are treated as the same, then
$n=1$ and player $i$'s strength is 1-dimensional. So,
$\btheta$ and $\bmu$ are 2-dimensional,
$p_{ij}=p_{ij}(\btheta)=e^{b\theta_i}/(e^{b\theta_i} + e^{b\theta_j})$,
and $-H^{-1}$ and $-H^{-1}J$ in Propositions \ref{prop1} and \ref{prop2}
simplify to
\begin{align}
\begin{split}
& -H^{-1}(\bmu)  = 
\begin{bmatrix}
\sigma_i^2 & 0\\
0 & \sigma_j^2
\end{bmatrix}
- w
\begin{bmatrix}
\phantom{1} \sigma_i^4 & -\sigma_i^2 \sigma_j^2\\
-\sigma_i^2\sigma_j^2 & \phantom{1} \sigma_j^4
\end{bmatrix}
,
\label{H-1}
\\
& -H^{-1}(\bmu) J(\bmu) = 
\begin{bmatrix}
k_{i}(s_{ij}-\hat{p}_{ij})\\
k_{j}(s_{ji}-\hat{p}_{ji})
\end{bmatrix}
,
\end{split}
\end{align}
where 
$\sigma_i^2$ and $\sigma_j^2$ are the variances of $\theta_i$ and $\theta_j$,
$\hat{p}_{ij}=p_{ij}(\bmu)$, 
$w=b^2 \hat{p}_{ij} \hat{p}_{ji} C$
and $k_i=b\sigma_i^2 C$,
with $C=(1+b^2 \hat{p}_{ij} \hat{p}_{ji} (\sigma_i^2 + \sigma_j^2))^{-1}$.
Now let 
\begin{subequations}
\begin{align}
& \hat{p}'_{ij}=p_{ij}(\bmu') = \frac{e^{b \mu'_i}}{e^{b \mu'_i}+e^{b \mu'_j}},
\label{p'}
\\
& C'=(1+b^2 \hat{p}'_{ij} \hat{p}'_{ji} (\sigma_i^2 + \sigma_j^2))^{-1},
\label{C'}
\\
& L_i = \hat{p}'_{ij}\hat{p}'_{ji}\sigma^2_{i}b^2C' = \frac{\hat{p}'_{ij} \hat{p}'_{ji} \sigma^{2}_{i} b^{2}}{1 + b^{2}\hat{p}'_{ij} \hat{p}'_{ji}(\sigma^{2}_{i} + \sigma^{2}_{j})}.
\label{Li}
\end{align}
\end{subequations}
Then, the diagonal elements of $-H^{-1}(\mu')$ can be rearranged, and we have the following update rules: 
\begin{subequations}
\begin{align}
& \mu^{'}_{i} = \mu_{i} + k_{i} (s_{ij}-\hat{p}_{ij}),
\label{mu'}
\\
& (\sigma^{2}_{i})' = \sigma^{2}_{i}(1-L_i).
\label{sig'}
\end{align}
\end{subequations}
If the variance is assumed to be the same for all players as in \cite{ing21}, 
then $\sigma_i=\sigma_j=\sigma$,
\begin{align}
k_i=k_j= \frac{b\sigma^2}{1+2b^2\sigma^2\hat{p}_{ij}\hat{p}_{ji}},
\label{kikj}
\end{align}
and \eqref{mu'} would be the same as equations (18) and (19) in \cite{ing21}.
There are two concerns with the update rules \eqref{mu'}-\eqref{sig'}.
First, with $L_i$ lying in (0, 1), the variance always decreases as time
goes on and may be overly small.
To illustrate this point, assuming that $i$ starts with initial $\sigma_i=200$.
Table \ref{sigmaK} reports the number of matches $i$ plays and 
the corresponding $L_i$, $\sigma'_i$ and $k_i$ values in 
\eqref{mu'}-\eqref{sig'}. For convenience, here we take
$\hat{p}'_{ij}=\hat{p}'_{ji}=0.5$, and 
Table \ref{sigmaK}(a) sets $\sigma_j=100$ for all players $i$ plays against,
while Table \ref{sigmaK}(b) uses $\sigma_j=200$.
Table \ref{sigmaK}(a) shows that, as the number of matches increases to 300 
and 500, the $L_i$ values decline to 0.003 and 0.002,
and the $\sigma_i$ values are about 20.76 and 16.12.
We found that though $L_i$ becomes rather small, $\sigma_i$ does not seem
stabilized yet.
We also found that after 50 and 150 matches, 
the $\sigma_i$ values are about 50 and 30,
and the corresponding $k_i$ values are about 12 and 4.5, 
which are exceedingly low comparing to the typical Elo's $K$ of 32.
The results in Table \ref{sigmaK}(b) are similar.
Actually, for the professional men's tennis data considered in 
Section \ref{data}, over 100 players have played more than 250 matches,
and some have even played more than 500 matches;
for instance, Novak Djokovic has played 685 matches, 
Rafael Nadal has 650 matches, and Roger Federer has 640 matches.
So, with the variance update formula
\eqref{sig'}, many of the $k_i$ values would become rather low 
and the prediction could be poor.
The second concern is that the underlying model for (20a)-(20b) makes a naive
assumption that players' strengths are time-invariant.
This assumption may not be realistic for real data, especially for
competitive young players.
Hereafter we refer to (20b) as the naive variance update rule.
We address these concerns by employing a strength evolution model similar to
Glicko and imposing some bounds on variance to prevent it from getting
overly high or low. Specifically,
we start with Glicko's strength evolution model \eqref{rw}, 
$\theta_{it}=\theta_{i,t-1}+e_{it}$.
Yet instead of using a constant variance $c^2$ for $e_{it}$ across players, 
we assume that the increase in uncertainty over time
may vary from player to player; that is,
$e_{it}$ follows $N(0, \eta_i^2)$. 
As an analogous to the variance update formula in \eqref{glicko},
now the player's variance will be increased by $\eta_i^2$, which gives
\begin{align}
(\sigma^{2}_{i})' = \sigma^{2}_{i}(1-L_i) + \eta_i^2.
\label{eta1}
\end{align}

We make a further assumption that the increase in uncertainty 
for more experienced players is likely to be smaller. As experienced
players tend to have played more games and hence smaller variances, we suggest
to take $\eta_i^2$ to be proportional to $\sigma_i^2$ or $\sigma_i^2 L_i$
so that $\eta_i^2$ would tend to be smaller for experienced players.
In what follows, we will examine the two settings: 
$\eta_i^2 = \alpha \sigma_i^2 L_i$ and $\eta_i^2 = \alpha \sigma_i^2$
with $\alpha \geq 0$. 
For $\eta_i^2 = \alpha \sigma_i^2 L_i$ with some $\alpha \geq 0$, 
\eqref{eta1} becomes
\begin{align}
(\sigma^{2}_{i})' 
= \sigma^{2}_{i}(1-AL_i),
\label{eta2}
\end{align}
where $A=1-\alpha$.
If $0 \leq \alpha \leq 1$, then $0 \leq A \leq 1$ 
and $A$ in \eqref{eta2} can be regarded as a reduction factor for variance 
adjustment. 
Note that a smaller $A$ corresponds to a lower reduction in the variance, 
$A=0$ means no reduction,
and $A=1$ if and only if $\eta_i^2=0$ in \eqref{eta1}, 
which means that the strength $\theta_i$ does not evolve over time.
Though the use of $A<1$ slows down the decay of $\sigma'$,
\eqref{eta2} still leads to $\sigma'_i < \sigma_i$.
So, we shall impose a lower bound $B$ on the standard deviation to prevent it from 
being overly low. The resulting update rule is
\begin{align}
(\sigma^{2}_{i})' = \max(B^2, \sigma^{2}_{i}(1- AL_i)).
\label{eta3}
\end{align}
In particular, $B=0$ means that no bound is imposed on $\sigma_{i}$.
For instance,
$(A, B)=(0, 0)$ corresponds to $(\sigma^{2}_{i})'=\sigma^{2}_{i}$,
which is the constant variance case;
and $(A, B)=(1, 0)$ gives the naive variance update rule \eqref{sig'}.
The equations \eqref{mu'} and \eqref{eta3} serve as our basic 
formulas for updating mean and variance. 
We present it in Algorithm \ref{algoLap}. 
Since \eqref{mu'} resembles the Elo formula, for convenience we refer to it 
as Elo and the variance incorporated formulas as vElo.
If $\alpha>1$, then $A=1-\alpha<0$ 
and $\sigma'_{i} > \sigma_i$, which is less likely
as it does not reflect the fact that the uncertainty about
a player's rating would become smaller as more matches are observed.
Alternatively, one may take $\eta_i^2 = \alpha \sigma_i^2$ for
some $\alpha \geq 0$ so \eqref{eta1} becomes
$(\sigma^{2}_{i})' = \sigma^{2}_{i}(1-L_i+\alpha)$; see 
Section \ref{subsec:eta} for more discussion about this option.


The variance update scheme \eqref{eta3} can also be incorporated into
the GenElo Surface model \citep{ing21}.
To start, if the current game is played on surface $m$,
the naive update formula \eqref{sig'} and 
the diagonal elements of $-H^{-1}$ in Proposition \ref{prop1} suggest that
\begin{align*}
& (\sigma_l^2)'
= \sigma_l^2 \left(1-b^2 \hat{p}_{ij} \hat{p}_{ji} \sigma_m^2 \rho_{ml}^2 C\right),\; l=1,...,n,
\end{align*}
where $C=(1+b^{2}\hat{p}_{ij}\hat{p}_{ji}(\sigma^{2}_{m}+ \tau^{2}_{m}))^{-1}$.
Then, we modify the above line 
by using a reduction factor $A$ and a lower bound $B$ as in \eqref{eta3}.
Finally, for the GenElo Surface model, 
Algorithm \ref{algoLapS0} presents the results of Theorem \ref{thm1},
which rewrites Ingram's formula \eqref{bmu'} in an interpretive way, 
and Algorithm \ref{algoLapS} presents the extension to incorporate variance update. We will refer to the variance incorporated one as vGenElo surface algorithm.

\begin{table} \footnotesize
\centering
\begin{tabular}{lrrrrrrrrr}
matches & 0 & 25 & 50 & 100 & 150 & 200 & 300 & 400 & 500
\\ 
$L_i$ & &  0.034 & 0.018 & 0.01 & 0.006 & 0.005 & 0.003 & 0.002 & 0.002
\\
$\sigma_i$ & 200 & 68.00 & 49.54 & 35.58 & 29.20 & 25.36 & 20.76 & 18.00 & 16.12
\\ 
$k_i$ & 138.5 & 23.7 & 12.8 & 6.7 & 4.5 & 3.4 & 2.3 & 1.7 & 1.4
\\
\end{tabular}\\
\smallskip
(a) $\hat{p}'_{ij}=\hat{p}'_{ji}=0.5$ and opponents' $\sigma_j=100$

\bigskip
\centering
\begin{tabular}{lrrrrrrrrr}
matches & 0 & 25 & 50 & 100 & 150 & 200 & 300 & 400 & 500
\\ 
$L_i$ & &  0.033 & 0.018 & 0.01 & 0.006 & 0.005 & 0.003 & 0.002 & 0.002
\\
$\sigma_i$ & 200 & 74.42 & 54.55 & 39.31 & 32.30 & 28.07 & 22.99 & 19.94 & 17.86
\\
$k_i$ & 138.5 & 23.1 & 12.6 & 6.6 & 4.5 & 3.4 & 2.3 & 1.7 & 1.4
\\
\end{tabular}\\
\smallskip
(b) $\hat{p}'_{ij}=\hat{p}'_{ji}=0.5$ and opponents' $\sigma_j=200$
\caption{Number of matches, $L_i$, $\sigma_i$ and $k_i$}
\label{sigmaK}
\end{table}

\begin{algorithm}[htbp]
	\caption{(vElo) Variance incorporated Elo system}
        \begin{enumerate}
 	\item Given a game outcome on surface $m$ by players 1 and 2.
	\item Given $0 \leq A \leq 1$, $B \geq 0$ and current skills of players 1 and 2: ($\mu_{i}, \sigma_{i}^{2}$), $i=1,2$.
	\item Let 
\begin{align*}
& b=\log(10)/400,\\
& \hat{p}_{12} = \frac{e^{b\mu_1}}{e^{b\mu_1}+e^{b\mu_2}}, \;\hat{p}_{21} = 1-\hat{p}_{12},\\
& C=(1+b^{2}\hat{p}_{12}\hat{p}_{21}(\sigma^2_{1} +\sigma^2_{2}))^{-1}.
\end{align*}
	\item For $i=1,2$,\newline
	(a) Let
		$$s_{ij} = 
		\left\{
			\begin{aligned}
			1	&\quad& \mbox{if } i \mbox{ wins},\\
			0	&\quad& \mbox{if } i \mbox{ loses}.
			\end{aligned}
		\right.$$ 
	(b) Let $k_i = \sigma_i^2 bC$. Update $\mu_{i}$ by
\begin{align*}
& \mu'_{i} = \mu_{i} + k_i(s_{ij} - \hat{p}_{ij})
\end{align*}
        (c) Let
\begin{align*}         
 & \hat{p}'_{12}= \frac{e^{b\mu'_{1}}}{e^{b\mu'_{1}} + e^{b\mu'_{2}}},\; \hat{p}'_{21}= 1-\hat{p}'_{12},\\
 & C'=(1+b^{2}\hat{p}'_{12}\hat{p}'_{21}(\sigma^2_{1} +\sigma^2_{2}))^{-1}
 \; {\rm and}\;
L_i = \hat{p}'_{12}\hat{p}'_{21}\sigma^2_{i}b^2C'.
\end{align*} 
Update $\sigma^{2}_{i}$ by
\begin{align*}
&(\sigma'_{i})^2 = \max(B^2, \sigma^{2}_{i}(1-AL_i)).
\end{align*}
        \end{enumerate}
\label{algoLap}
\end{algorithm}

\begin{algorithm}[htbp]
	\caption{(GenElo Surface) A new expression for update rule \eqref{bmu'} by Ingram}
	\begin{enumerate}
	\item Given constant variances $\sigma_l^{2}$, for $l=1,2,3$.\newline
        Given correlation coefficients $\rho_{q l}$, for $q,l$=1, 2, 3.\newline
	Given a game outcome on surface $m$ by players 1 and 2. 
        \item Given current skills of players 1 and 2: $\bmu_{i}, i =1,2$, where $\bmu_{i}=(\mu_{i1}, \mu_{i2}, \mu_{i3})$.
	\item Let \newline
       (a) $b=\log(10)/400$, $\hat{p}_{12} = e^{b\mu_{1m}}/(e^{b\mu_{1m}}+e^{b\mu_{2m}})$, $\hat{p}_{21} = 1-\hat{p}_{12}$, \newline
       (b) $C=(1+2b^{2}\hat{p}_{12}\hat{p}_{21}\sigma_m^{2})^{-1}$.
	\item For $i$ = 1, 2,\newline
	(a) Let
		$$s_{ij} = 
		\left\{
			\begin{aligned}
			1	&\quad& \mbox{if } i \mbox{ wins},\\
			0	&\quad& \mbox{if } i \mbox{ loses}.
			\end{aligned}
		\right.$$ 
	(b) For surface $l$ = 1, 2, 3, let
		$$\mu_{il}^{'} = 
		\left\{
                 \begin{array}{ll}
		\mu_{il} + \sigma_m^{2}bC(s_{ij}-\hat{p}_{ij}) & \mbox{if } l = m,\\
		\mu_{il} + \sigma_m \sigma_l \rho_{ml}bC(s_{ij}-\hat{p}_{ij}) & \mbox{if } l \neq m.
	        \end{array} \right.$$
	\end{enumerate}
\label{algoLapS0}
\end{algorithm}

\begin{algorithm}
	\caption{\footnotesize(vGenElo Surface) Variance-incorporated update rules for GenElo Surface model}
The procedure is the same as Algorithm \ref{algoLapS0} except Steps 2, 3(b), and 4(b)(c):
\begin{enumerate}
\item[2.]
 Given $0 \leq A \leq 1$, $B \geq 0$ and current skills of players 1 and 2: ($\bmu_{i}, \bsigma_{i}^{2}$), $i=1,2$, where $\bmu_{i} = (\mu_{i1}, \mu_{i2}, \mu_{i3})$ and $\bsigma^{2}_{i} = (\sigma^{2}_{i1},\sigma^{2}_{i2},\sigma^{2}_{i3})$.
\item[3(b).] Let 
$C=(1+b^{2}\hat{p}_{12}\hat{p}_{21}(\sigma^2_{1m} +\sigma^2_{2m}))^{-1}$
\item[4(b).] For surface $l =1,2,3$, let
\begin{align*}
& k_{il} = \sigma_{il}\sigma_{im}\rho_{m l}bC;
\end{align*}
	update $\mu_{il}$ by
\begin{align*}
& \mu'_{il} = \mu_{il} + k_{il}(s_{ij} - \hat{p}_{ij}).
\end{align*}
\item[ (c).] 
For surface $l =1,2,3$, let
\begin{align*}
& C'=(1+b^{2}\hat{p}'_{12}\hat{p}'_{21}(\sigma^2_{1m} +\sigma^2_{2m}))^{-1},\\
& L_{il} = \hat{p}'_{12}\hat{p}'_{21}\sigma^{2}_{im}\rho^{2}_{m l}b^{2}C';
\end{align*}
	update $\sigma^{2}_{il}$ by
\begin{align*}
& (\sigma'_{il})^2 = \max(B^2, \sigma^{2}_{il}(1-AL_{il})).
\end{align*}
\end{enumerate}
\label{algoLapS}
\end{algorithm}

\section{Experiments}
We conduct numerical studies to evaluate the accuracy of the single Newton-Raphson step in \eqref{S'}, see
Appendix \ref{append:S'}. The results are quite satisfactory;
In what follows we will evaluate the performance of our algorithms on men's tennis matches.

\subsection{Dataset}
\label{data}
We obtain professional men's tennis matches during 2010 to 2019 from 
\url{https://github.com/JeffSackmann/tennis_atp} by Sackmann. 
We remove retirements, defaults, and walkovers, and
drop matches played on carpet as this surface is rarely used recently.
The remaining data has 771 players in total; 
among them, 135  players have more than 250 appearances.
Matches from 2010-2017 are used as the train set and those
from 2018-2019 as the test set. 
There are 20,437 matches in the train set and 5,100 matches in the test set.
The train set is used to estimate model parameters and the test set is to evaluate the prediction accuracy for the test set.
The above data preparation is the same as \cite{ing21} except that here
the data source is different and
the playing surfaces are classified into three categories rather than four.
The data summary is in Table \ref{tab:data}.

\begin{table} \footnotesize
\centering
\begin{tabular}{lrrr}
 &Train	&Test	&Total \\ 
Number of matches &	20437	&5100	&25537\\
Hard	&11613	&2907	&14520\\
Clay	&6399	&1559	&7998\\
Grass	&2425	&634	&3059
\end{tabular}
\caption{Data summary}
\label{tab:data}
\end{table}

\subsection{Parameter settings}
\label{sec:para}
The initial rating of each player is set as 1500.
For Algorithm \ref{algoLap}, the playing surface is not taken into account
and there is just one initial variance $\sigma^2$.
For Algorithm \ref{algoLapS0} and Algorithm \ref{algoLapS}, 
with three playing surfaces, 
there are three initial variances associated with the three surfaces
and three correlation coefficients between surfaces.
These model parameters are to be estimated
by minimizing the negative log-likelihood function over the train set, 
as suggested by \cite{Glicko} and \cite{ing21}.
When the variance update is to be incorporated, the parameters $A$ and $B$ also
need to be inferred. In our experiments, we consider a selection of $(A, B)$
values, and present the best models associated with each combination of $(A, B)$.


\subsection{Results}
\label{sec:res}
This section consists of three parts.
Section \ref{subsec:res} provides general forecasting
evaluations of the algorithms, 
Section \ref{subsec:new-player} presents results for new
players, and Section \ref{subsec:eta} compares the use of different
$\eta_i^2$ in \eqref{eta1}.

\subsubsection{General evaluations}
\label{subsec:res}
Table \ref{nRho} reports results based on Algorithm \ref{algoLap} and 
various choices of $(A, B)$. 
Recall that $(A, B)=(0, 0)$ gives an Elo-type algorithm, which 
assumes constant variance ($\sigma_i=\sigma$ for all $i$)
and updates only $\mu_i$ in step 4(b).
In Table \ref{nRho}(a), the left panel is based on $(A, B)=(0, 0)$;
the right panel is for $(A, B)=(1, 0)$, which is equivalent
to the update rules \eqref{mu'}-\eqref{sig'},
where \eqref{sig'} is the naive variance update formula based on negative
inverse of Hessian.
For each of the two settings, we conduct experiments with selected
initial $\sigma$ values ranging between 50 and 200, and
the results associated with the smallest negative log-likelihood 
value for the train set are reported.
The column $\sigma$ stands for the initial $\sigma$ values. The next two
columns report the prediction accuracy for the test set, and
the average negative log-likelihood based on the train set. 
A smaller negative log-likelihood value means a better model. 
We found from Table \ref{nRho}(a) that the constant variance setting
outperforms the naive variance update in terms of prediction accuracy.
This can be expected because the $\sigma$ values could become exceedingly low
as is reported in Table \ref{sigmaK}.
For the constant variance setting, the best model occurs at 
$\sigma=80$ and the associated accuracy rate is 0.6338.
By \eqref{kikj}, for a match with an equal win probability 0.5,
$\sigma=80$ roughly corresponds to $K = 33.3$ in Elo's formula.
For the naive variance update, the prediction accuracy 
is found to be 0.6199, which occurs when the initial $\sigma$ is 200.

Next, Table \ref{nRho}(b) sets $(A, B)=(1, 75)$ and (1, 80),
where $A$=1 means that $\theta$ does not evolve over time. 
Here the bound $B$ is chosen to be around 80, a value suggested by the best 
$\sigma$ in the left panel of Table \ref{nRho}(a).
For each of the two settings of $(A, B)$, 
we conduct experiments with a selected 
initial $\sigma$ values ranging between 50 and 200.
The results show that the best models occur when the initial $\sigma$ 
values are between 120 and 130, and the associated average negative 
log-likelihood values are smaller than those in Table \ref{nRho}(a).
The corresponding accuracy rates are 0.6365 and 0.6350, 
which slightly improves the constant variance setting in Table \ref{nRho}(a).
Then, Table \ref{nRho}(c) sets $B$=80 and considers
time-varying strengths by taking $A=1/3$ and 1/5. The 
accuracy rates are 0.6381 and 0.6387, indicating that the
use of dynamic models further enhances the prediction accuracy;
also, the average negative log-likelihood values have further decreased.
To sum up, Table \ref{nRho} shows that the constant variance setting
outperforms the naive variance update \eqref{sig'},
but the accuracy gains with variance update increase substantially 
if a suitable bound and time-varying strengths are employed.

\begin{table} \footnotesize
\centering
\begin{tabular}{lrrrrrlrrr}
$\sigma$ & accuracy & neg-loglike &&& $\sigma$ & accuracy & neg-loglike\\
80 & 0.6338 & 0.5958 & & & 200 & 0.6199 & 0.6006
\end{tabular}\\
\smallskip
(a) Left: $(A, B)=(0, 0)$ (only update $\mu$); Right: $(A, B)=(1, 0)$.

\bigskip
\centering
\begin{tabular}{lrrrrrlrrr}
$\sigma$ & accuracy & neg-loglike &&& $\sigma$ & accuracy & neg-loglike\\
130 & 0.6365 & 0.5952 & & & 120 & 0.6350 & 0.5955
\end{tabular}\\
\smallskip
(b) Left: $(A, B)=(1, 75)$; Right: $(A, B)=(1, 80)$.

\bigskip
\centering
\begin{tabular}{lrrrrrlrrr}
$\sigma$ & accuracy & neg-loglike &&& $\sigma$ & accuracy & neg-loglike\\
120 & 0.6381 & 0.5950 &&& 110 & 0.6387 & 0.5950
\end{tabular}\\
\smallskip
(c) Left: $(A, B)=(1/3, 80)$; Right: $(A, B)=(1/5, 80)$.
\caption{Prediction accuracy for test data by Algorithm \ref{algoLap} (Elo v.s. vElo)}
\label{nRho}
\end{table}

\begin{figure}
\begin{tabular}{cc}
  \includegraphics[width=0.45\linewidth,height=5cm]{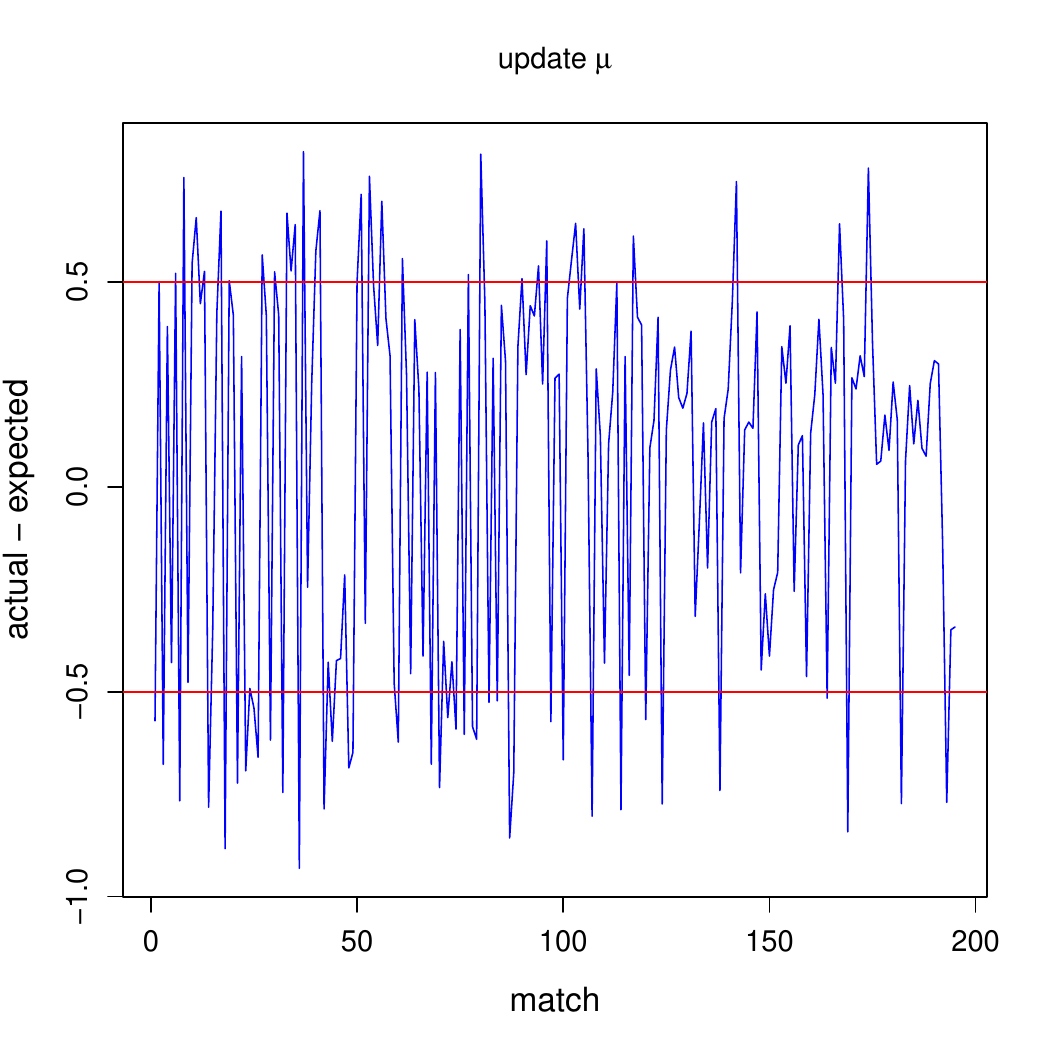}
&
  \includegraphics[width=0.45\linewidth,height=5cm]{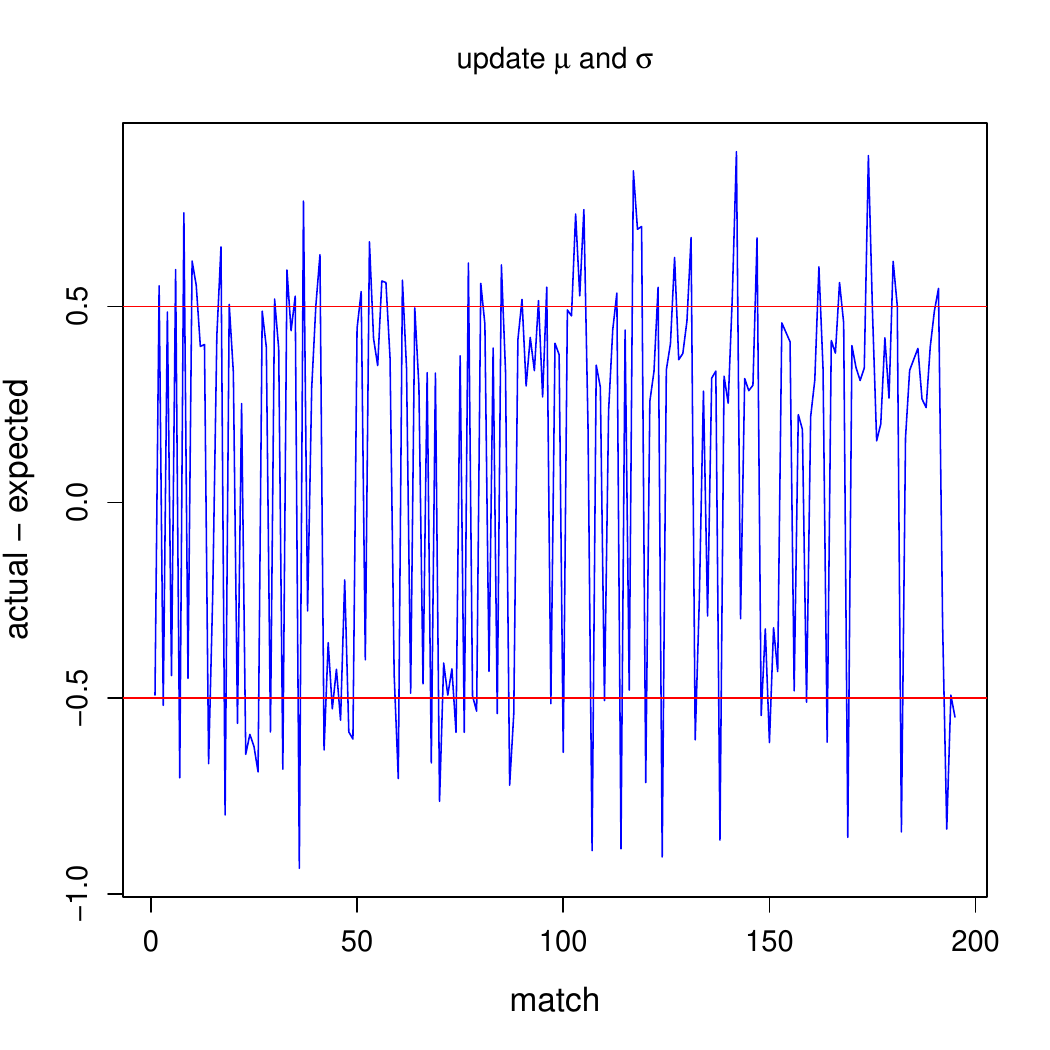}
\end{tabular}
\caption{Differences between actual and expected outcomes for Medvedev.}
\label{medve1}
\end{figure}

\begin{table} \footnotesize
\centering
\begin{tabular}{rrrrrrrr}
$\sigma_{clay}$ & $\sigma_{grass}$ & $\sigma_{hard}$ & $\rho_{cg}$ & $\rho_{ch}$ & $\rho_{gh}$ & accuracy & neg-loglike\\
91.62 & 98.71 & 80.37 & 0.47 & 0.72 & 0.84 & 0.6452 & 0.5910
\end{tabular}\\
\smallskip
(a) results by Algorithm \ref{algoLapS0} (GenElo Surface model)

\bigskip
\centering
\begin{tabular}{rrrrrrrrrr}
A& B & $\sigma_{clay}$ & $\sigma_{grass}$ & $\sigma_{hard}$ & $\rho_{cg}$ & $\rho_{ch}$ & $\rho_{gh}$ & accuracy & neg-loglike\\
1/4 & 0 & 130.43 & 142.15 & 120.98 & 0.44 & 0.70 & 0.83 & 0.6487 & 0.5885\\
1/5 & 0 & 124.38 & 134.85 & 114.14 & 0.45 & 0.71 & 0.83 & 0.6481 & 0.5885\\
1/4 & 80 & 128.48 & 142.47 & 119.55 & 0.46 & 0.71 & 0.83 & 0.6469 & 0.5895\\
1/5 & 80 & 124.09 & 134.28 & 114.36 & 0.47 & 0.71 & 0.84 & 0.6473 & 0.5896
\end{tabular}\\
\smallskip
(b) results by Algorithm \ref{algoLapS} (vGenElo Surface model) with various ($A$,$B$)
\caption{Prediction accuracy for test data by Algorithm \ref{algoLapS0} and \ref{algoLapS}}
\label{Rho}
\end{table}

Table \ref{Rho}(a) reports results by Algorithm \ref{algoLapS0},
which is a re-expression of the update rules
\eqref{bmu'} for Ingram's GenElo Surface model.
\cite{ing21} assumes that the covariance $\Sigma$ of the strength $\btheta$ 
is the same across players and over time.
The matrix $\Sigma$ involves 
$(\sigma_{clay}, \sigma_{grass}, \sigma_{hard}, 
\rho_{cg}, \rho_{ch}, \rho_{gh})$, where
$\rho_{cg}$, $\rho_{ch}$, $\rho_{gh}$ are the correlations between 
clay and grass, clay and hard, and grass and hard, respectively.
These parameters are estimated by minimizing the negative log-likelihood
function over the train data. 
As in \cite{ing21}, we found that grass has the highest estimated
standard deviation, followed by clay, and then hard courts; 
and the estimated correlation coefficients are all positive, with the
correlation between grass and hard being the greatest,
and the correlation between clay and grass being the smallest.
These values will be fixed for the test data.
The last two columns report the the prediction accuracy for the test data
and the average negative log-likelihood based on the train data.
It shows that, with the incorporation of playing surface, the prediction 
accuracy is 0.6452, which improves over the accuracy of 0.6338 in the left 
panel of Table \ref{nRho}(a). This agrees with the findings in \cite{ing21}.

Finally, Table \ref{Rho}(b) presents results by Algorithm \ref{algoLapS},
which corresponds to vGenElo Surface, the variance incorporated GenElo Surface model.
Consider a selection of the reduction factor $A \in \{1, 1/2, 1/3, 1/4, 1/5\}$ 
and the lower bound $B \in \{0, 80, 100\}$. 
For each of the 15 combinations of $(A, B)$, we obtain the estimates of
correlations and initial standard deviations
by minimizing the negative log-likelihood values based on the train data.
The results for $(A, B)=(1/4, 0)$, (1/5, 0), (1/4, 80), (1/5, 80) are reported;
the rest are similar and omitted.
We found that
the estimates of initial standard deviations here are all larger than those 
in Table \ref{Rho}(a),
and that the estimates for $A=1/4$ are larger than those for $A=1/5$. 
This can be explained by the facts that the variance update scheme 
\eqref{eta3} usually reduces the variances, and the use of higher reductions 
could allow a larger initial standard deviation.
Table \ref{Rho}(b) also showed that, with the use of variance update,
the average negative log-likelihood values are smaller
and the prediction accuracy slightly increases. Recall in Section \ref{LapDyn} that $B$ is the lower bound on $\sigma$ to prevent it from being overly low, and that $B=0$ corresponds to not imposing lower bound on $\sigma$. In contrast to Table \ref{nRho}(a) where $B=0$ gives poor prediction accuracy, Table \ref{Rho}(b) shows that $B=0$ performs well. This may due to the reason that when matches are classified into clay, grass and hard, there would be less matches in each playing surface, and hence the updated $\sigma_{clay}$, $\sigma_{grass}$ and $\sigma_{hard}$ have not declined too much.

We note that in Table \ref{nRho} 
the accuracy gain from using variance update is from 0.6338 to 0.6381 or even 0.6387. 
With 5,100 matches in test data, this gain is due to around 22 to 25 matches, 
obtained by (0.6381-0.6338)*5,100=21.93 and (0.6387-0.6338)*5,100=24.99.
However, when the playing surface is taken into account,
Table \ref{Rho} shows that the increases in forecast accuracy from
0.6452 to 0.6487 is indeed very small;
with 5,100 matches in test data, this is due to about 17 to 18 matches,
found by (0.6487-0.6452)*5,100=17.85.
These observations suggest that to some extent the uncertainty in players'
strengths may be attributed to the playing surface. 
Therefore, to further investigate the advantage of variance update
without the effect of playing surface,
we conduct experiments for hard-surface matches. 
We choose hard surface as it makes up about 57\% of the data, 
found by 14,520/25,537 = 0.568.
As shown in Table \ref{tab:data}, for hard-surface games,
there are 11,613 matches in the train data
and 2,907 matches in the test set.
Table \ref{tab:hard} shows that the prediction accuracy with the use of 
variance update scheme has increased from 0.6462 to 0.6535.

\begin{table} \footnotesize
\centering
\begin{tabular}{lrr}
$\sigma$&    accuracy&    neg-loglike\\
     85 & 0.6462 & 0.5978
     \end{tabular}\\
\smallskip
(a) results by Algorithm \ref{algoLap} with only $\mu$-update

\bigskip
\centering
\begin{tabular}{lrrr}
$\sigma$&    accuracy&    neg-loglike&    (A,B)\\
    120 & 0.6535 & 0.5949 & (1/4, 50)\\
    110 & 0.6531 & 0.5951 & (1/5, 50)\\
    120 & 0.6514 & 0.5950 & (1/4, 60)\\
    115 & 0.6511 & 0.5951 & (1/5, 60)
\end{tabular}\\
\smallskip
(b) results by Algorithm \ref{algoLap} with factor $A$ and bound $B$
\caption{Prediction accuracy for hard-surface matches in test data by Algorithm \ref{algoLap}}
\label{tab:hard}
\end{table}

To investigate whether the improvements in prediction accuracy are meaningful,
one possibility is to obtain an uncertainty estimate by bootstrapping. 
The bootstrap method is a useful resampling technique for obtaining uncertainty estimates. 
For independent and identically distributed observations of size $n$, 
it constructs a number of resamples (of size $n$) with replacement 
from the observed data set. 
However, since the sport data are collected over time and the algorithms
considered here heavily depend on the order of data,
the bootstrap samples may not be appropriate because 
the data ordering is disrupted.

Alternatively, one may conduct hypothesis testing to test
whether the use of variance update has increased the prediction accuracy. 
As the accuracy rates are evaluated on the same testing data,
we shall conduct McNemar test for dependent proportions.
To illustrate the test, suppose that the prediction performance of two 
classifiers are to be compared.
Let $\pi_i$, $i=1,2$, be the probability of correct prediction 
with classifier $i$;
and let $n_{ij}$ be the number of cases for which classifier $i$
predicts correctly and classifier $j$ predicts incorrectly.
Then, for testing $H_0: \pi_1 \geq \pi_2$ against $H_1: \pi_1 < \pi_2$, 
the McNemar test statistic is
$Z= (n_{21}-n_{12})/\sqrt{n_{12}+n_{21}}$,
and $H_0$ is rejected if $Z > z_{\alpha}$,
where $\alpha$ is the significance level and $z_{\alpha}$ is the upper
$\alpha$ quantile of the standard normal distribution;
see \citet[Chapter 10]{agresti02} for a detailed account of the test.
Let classifier 1 correspond to the formula with only mean update, 
where the prediction results are in Tables \ref{nRho}(a), \ref{Rho}(a)
and \ref{tab:hard}(a);
and let classifier 2 correspond to the formula
with both mean and variance update,
where the prediction results are in Tables \ref{nRho}(b)(c), \ref{Rho}(b)
and \ref{tab:hard}(b).
The latter involves different $(A, B)$ values, and here
we choose the value associated with the smallest negative log-likelihood.
Table \ref{tab:mcN} reports McNemar test statistics and $p$-values.
The first two rows are based on Table \ref{nRho}, where the $p$-values are
around 0.002 to 0.004, indicating that the use of variance update has
significantly improved the prediction accuracy.
The $p$-value in the third row is 0.1772, 
which agrees with our earlier finding that, when the playing surface
is incorporated into the model, the increase in accuracy
from 0.6253 to 0.6487 is small and insignificant.
The last row shows a $p$-value of 0.051 
when only hard-surface matches 
are considered, suggesting that the variance update scheme could be useful.
We remark here that a larger increase in accuracy does not guarantee a smaller
$p$-value. For instance, 
in Table \ref{tab:hard}(b) for hard-surface matches,
though the accuracy gain for $(A, B)=(1/5, 50)$ 
is smaller than that for $(A, B)=(1/4, 50)$,
the McNemar test gives a $p$-value of 0.0478, which is slightly smaller
than 0.0510.

\begin{table} \footnotesize
\centering
\begin{tabular}{llll}
classifier 1 &   classifier 2 &   $Z$ &   $p$-value\\
   Table \ref{nRho}(a)  & Table \ref{nRho}(c) with $(A,B)=(1/3, 80)$ & 2.668 & 0.0038\\
   Table \ref{nRho}(a)  & Table \ref{nRho}(c) with $(A,B)=(1/5, 80)$ & 2.887 & 0.0019\\
   Table \ref{Rho}(a)  & Table \ref{Rho}(b) with $(A,B)=(1/4, 80)$ & 0.926 & 0.1772\\
   Table \ref{tab:hard}(a)  & Table \ref{tab:hard}(b) with $(A,B)=(1/4, 50)$ & 1.635 & 0.0510
\end{tabular}
\caption{McNemar test for $H_0: \pi_1 \geq \pi_2$ v.s. $H_1: \pi_1 < \pi_2$}
\label{tab:mcN}
\end{table}

\subsubsection{Evaluation of new players}
\label{subsec:new-player}
It has been noted that the competitive ability for young players may 
improve in sudden bursts; see \cite{simon97} and \cite{glick01}.
First, for individual players, we consider Daniil Medvedev.
Medvedev has a total of 195 appearances in this dataset,
with his first appearance in 2016 at the age of 20. 
Figure \ref{medve1} presents the difference between the actual outcome
and the expected outcome, $(s - \hat{p})$. The expected outcome $\hat{p}$
are based on the two settings of Table \ref{nRho}(a); that is,
the left plot of Figure \ref{medve1} is under the constant variance setting 
with $\sigma=80$,
and the right plot is under the naive variance update with initial
$\sigma=200$.
Note that a value of $(s - \hat{p})$ within (-.5, .5) gives a correct 
prediction, $(s - \hat{p}) > .5$ indicates an underestimate of player's strength,
and $(s - \hat{p}) < -.5$ corresponds to an overestimate of player's strength.
These plots show that, for the first half of the 195 matches, 
the constant variance setting tends to have larger 
$|s - \hat{p}|$ values than the naive variance update, indicating that the 
use of variance update helps to capture the player's strength in early stage.
However, it also shows that the performance of variance update deteriorates 
in the second half of the 195 matches.
As explained in Section \ref{LapDyn}, this may possibly due to 
over-reducing player's $\sigma$,
and it can be addressed by employing a dynamic model together with 
the constraint of a lower bound on the variance.

Next, we investigate the performance of {\it new} players.
We consider new players to be those who have not
appeared in the first $N$ matches of the data,
and collect the first $n$ matches of each new player in the data.
Presumably most of the new players are young.
For each new player $i$, we compute 
the prediction accuracy with and without variance update, 
denoted as $acc_i^{(1)}$ and $acc_i^{(0)}$, respectively,
based on this player's first $n$ matches.
Then, 
let $m_{01}$ denote the number of new players where the use of 
constant variance outperforms that with variance update; 
that is, 
$m_{01}=|\{\mathrm{new \, player}\, i: acc_i^{(0)} > acc_i^{(1)}\}|$. 
Similarly, 
let $m_{10}=|\{\mathrm{new \, player}\, i: acc_i^{(1)} > acc_i^{(0)}\}|$
and $m_{00}=|\{\mathrm{new \, player}\,i: acc_i^{(0)} = acc_i^{(1)}\}|$.
In our experiments, we set $N=5,000$ and $n=20, 30, 40$.
We found that, among the 771 players in the whole data set, 
670 are considered as new players.
Table \ref{tab:firstN} reports results by 
Algorithm \ref{algoLap} with $(A, B)=(1/5, 80)$.
The left panel is based on the whole data.
It shows that 93 out of the 670 new players
have played more than 20 matches;
and among these 93 players, the constant variance setting achieves higher
prediction accuracy for 15 players, 
the variance update scheme works better for 23 players,
and the two settings have the same accuracy rates for 55 players. 
In general, the results show that $m_{10}$ is larger than $m_{01}$; 
that is, for a majority of new players, the prediction accuracy with variance update is higher.
The right panel of Table \ref{tab:firstN} is based on hard-surface matches only.
It also demonstrates that $m_{10}$ is greater than $m_{01}$.

\begin{table} \footnotesize
\centering
\begin{tabular}{lrrrrrrlrrrr}
$n$ &  $m_{01}$  & $m_{10}$ & $m_{00}$ & total& & & $n$ & $m_{01}$  & $m_{10}$   & $m_{00}$ & total\\
20 &  15 &  23 & 55 &93 & & & 20 &  10 & 15 & 30& 55\\
30 &  12 &  28 & 29 &69 & & & 30 &  9 &  13 & 17& 39\\
40 &  13 &  28 & 23 &64 & & & 40 &  8 &  11 & 15& 34
\end{tabular}\\
\smallskip
Left: Using the whole data; Right: Using only the hard-surface matches.
\caption{Prediction based on the first $n$ matches of new players.
($m_{01}=|\{i: acc_i^{(0)} > acc_i^{(1)}\}|$, 
$m_{10}=|\{i: acc_i^{(1)} > acc_i^{(0)}\}|$
and $m_{00}=|\{i: acc_i^{(0)} = acc_i^{(1)}\}|$, 
where $i$ ranges over new players.)}
\label{tab:firstN}
\end{table}

\subsubsection{Choices of $\eta_i^2$}
\label{subsec:eta}
It is not necessary to restrict $\eta_i^2$ in \eqref{eta1} to the form 
$\alpha \sigma_i^2 L_i$. 
For instance, we may take $\eta_i^2=\alpha \sigma_i^2$ or
take $\eta_i^2$ as a constant. 
Below we will assess the performance with these two choices
of $\eta_i^2$.
First, consider $\eta_i^2=\alpha \sigma_i^2$ for some $\alpha>0$.
For convenience, write
$$(\eta_i^{(1)})^2=\alpha^{(1)} \sigma_i^2 L_i
\quad {\rm and} \quad
(\eta_i^{(2)})^2=\alpha^{(2)} \sigma_i^2.$$ 
Equating the above two equations gives
$\alpha^{(1)} L_i = \alpha^{(2)}$, where $L_i$ is defined in \eqref{Li}.
So, from theoretical perspective, $\eta_{i}^{(1)}$ and $\eta_{i}^{(2)}$ are likely to give similar results when
$\alpha^{(1)} L_i \approx \alpha^{(2)}$.
Suitable $\alpha^{(2)}$ values can be implied from 
Table \ref{nRho}(c).
To start, in this table we have $A=1/3, 1/5$, which correspond to $\alpha^{(1)}=1-A=2/3, 4/5$.
Further, the initial $\sigma$ is about 120 and the lower bound $B$ on 
$\sigma$ is set to be around 80;
with $\hat{p}_{ij}=$0.1, 0.3, 0.5 and ($\sigma_{1}, \sigma_{2}$) = (120, 120), (120, 80), (80, 120), (80, 80), the $L_{i}$ values computed from \eqref{Li}
are found to be between around 0.02 and 0.1.
Table \ref{tab:Li} reports the $L_{i}$ values.
Therefore, by the relation $\alpha^{(2)}=L_i \alpha^{(1)}$,
the $\alpha^{(2)}$ value in $\eta_{i}^{(2)}$ may be chosen to be between
about 0.01 and 0.08. We run experiments with $B$ = 75, 80, and
$\alpha^{(2)}$ = 0.01, 0.02,..., 0.08. 
Table \ref{tab:eta}(a) shows that the smallest
negative log-likelihood value occurs at around $\alpha^{(2)}=0.03$
with $B=75$ and initial $\sigma=130$, and the accuracy rate is 0.6369. The results for $\alpha^{(2)}$ = 0.06, 0.07 and 0.08 are not reported as the negative log-likelihood values are much larger.

\begin{table} \footnotesize
\centering
\begin{tabular}{lrrrr}
           &    \multicolumn{4}{c}{$(\sigma_i, \sigma_j)$}\\
$\hat{p}'_{ij}$ & (120, 120) & (120, 80) & (80, 120) & (80, 80)\\
0.5 &  0.096 & 0.102 & 0.045 & 0.048\\
0.3 &  0.083 & 0.088 & 0.039 & 0.041\\
0.1 &  0.040 & 0.040 & 0.018 & 0.018
\end{tabular}
\caption{$L_i$ values computed from \eqref{Li} using various $(\hat{p}'_{ij}, \sigma_{i}, \sigma_j)$}
\label{tab:Li}
\end{table}

Next, consider the case where $\eta_i^2$ is a constant.
Although this would make the method more similar to Glicko, it would still be distinct, as we are updating from match to match rather than from one rating
period to another, and also the inference procedure is different.
To assess the performance of Algorithm \ref{algoLap} with
$(\sigma'_i)^2=\mbox{max}(B^2, \sigma_i^2(1-L_i)+\eta^2)$, 
a suitable range of $\eta$ should be chosen.
As described in the previous paragraph, Table \ref{nRho}(c) suggests that 
$\alpha = 2/3, 4/5$, the initial $\sigma$ is about 120
and the bound $B$ is about 80.
By Table \ref{sigmaK}, the $L_{i}$ value is about 0.033 if $\sigma$ is between around 80 and 120.
Therefore, setting $\eta^{2} = \alpha \sigma^{2}_{i}L_{i}$ gives
that $\eta$ is between about 12 and 20.
In the experiment, we consider $\eta \in \{10, 12, 15, 20, 25\}$;
for each $\eta$, we try a variety of
initial $\sigma$ values ranging between 50 and 200, and report the
results associated with the smallest negative log-likelihood value for the 
train set. Table \ref{tab:eta}(b) shows that the smallest negative log-likelihood values occur at $B=75$ and $(\eta, \sigma)=(12, 130)$ and (15, 125), and the corresponding accuracy rates are 0.6369 and 0.6363.

Overall, the prediction accuracy of the above two settings of $\eta_{i}^{2}$ are just slightly lower than and comparable to that in Table \ref{nRho}(c).
A possible advantage of $\eta_{i}^{2} = \alpha \sigma_{i}^{2} L_{i}$ is the inclusion of the $L_{i }$ term. As can be seen from Table \ref{tab:Li}, for fixed ($\sigma_{i}$, $\sigma_{j}$), the largest $L_{i}$ value occurs at $\hat{p}'_{ij}$ = 0.5, 
corresponding to the situation where the two players are well-matched in strength,
and the $L_{i}$ value declines when the gap in strength gets wide. 
The interpretation is that the variance addition $\eta_i^2$ depends on the disparity in strength.
 In general, the wider the disparity, the more certain the game outcome, and hence the smaller the variance addition.

\begin{table} \footnotesize
\centering
% ...
\begin{tabular}{lrrrrrlrrr}
$\alpha^{(2)}$ & $\sigma$ & accuracy & neg-loglike &&& $\alpha^{(2)}$ & $\sigma$ & accuracy & neg-loglike\\
0.01 & 130 & 0.6367 & 0.5951 &&& 0.01 & 125 & 0.6348 & 0.5955\\
0.02 & 130 & 0.6371 & 0.5948 &&& 0.02 & 125 & 0.6353 & 0.5953\\
0.03 & 130 & 0.6369 & 0.5945 &&& 0.03 & 130 & 0.6363 & 0.5951\\
0.04 & 115 & 0.6383 & 0.5952 &&& 0.04 & 120 & 0.6373 & 0.5953\\
0.05 & 105 & 0.6391 & 0.5973 &&& 0.05 & 120 & 0.6391 & 0.5973
\end{tabular}\\
\smallskip
Left: $B= 75$ \hspace{4cm} Right: $B= 80$\\
(a) Using $\eta_i^2=\alpha^{(2)} \sigma^{2}_{i}$ in \eqref{eta1}

\bigskip
\centering
\begin{tabular}{lrrrrrlrrr}
$\eta$ & $\sigma$ & accuracy & neg-loglike &&& $\eta$ & $\sigma$ & accuracy & neg-loglike\\
10 & 130 & 0.6369 & 0.5950 &&& 10 & 120 & 0.6352 & 0.5955\\
12 & 130 & 0.6369 & 0.5949 &&& 12 & 125 & 0.6348 & 0.5954\\
15 & 125 & 0.6363 & 0.5949 &&& 15 & 125 & 0.6357 & 0.5953\\
20 & 110 & 0.6361 & 0.5968 &&& 20 & 115 & 0.6363 & 0.5968\\
25 &  95 & 0.6381 & 0.5996 &&& 25 & 100 & 0.6377 & 0.5996
\end{tabular}\\
\smallskip
Left: $B= 75$ \hspace{4cm} Right: $B= 80$\\
(b) Using $\eta_i^2 = \eta^2$ in \eqref{eta1}
\caption{Prediction accuracy with different choices of $\eta_i^2$ in \eqref{eta1}}
\label{tab:eta}
\end{table}

\section{Conclusions}
This paper describes rating algorithms based on the
Laplace approximation for posterior distribution, together with the 
dynamic modeling for players' strengths.
We have shown how variance estimation can be incorporated into
the Elo system as well as GenElo surface model \citet{ing21}.
We have also rewritten Ingram's update formula for
the GenElo surface model in a simpler form.
Our algorithms differ from Glicko in two ways.
One is that our algorithm updates parameters after each game, 
while Glicko updates parameters after observing game
data over a rating period, where the number of games per player is 
suggested to be around 5-10 in a rating period \citep{Glicko}.
The other is that the variance addition $\eta_i^2$ in \eqref{eta1} can be regarded as 
an analogous to the constant $c^2$ of Glicko's update rule in \eqref{glicko}, 
but $\eta_i^2$ is assumed to vary among players and be time-dependent.
For the experiments on men's professional matches,
we have found that the uncertainty in a player's
strength may largely be attributed to the playing surface,
and that our variance update scheme improve the prediction accuracy when only hard-surface games are considered. 
We have also observed that the variance update may better reflect the volatile
strengths of new players. 
It would be interesting to incorporate the additional information 
such as margins of victory into our proposed framework.
We leave it as future work.

\appendix
\section{Proof of Proposition \ref{prop1}}
\label{append:prop1}
We need the following lemma.
\begin{lemma}
Let $\ell(\btheta)$ be the logarithm of posterior distribution as in \eqref{t},
and $H$ be the Hessian as in \eqref{JH}.
Then, $-H^{-1}=\Sigma (I- D \Sigma)^{-1}$. 
\label{invH}
\end{lemma}

\Proof
The result can be obtained by writing
\begin{align}
H^{-1} & =(-\Sigma^{-1}+ D)^{-1} 
\nonumber\\
&= -\Sigma -\Sigma(I-D \Sigma)^{-1} D \Sigma
\nonumber\\
&= -\Sigma \left(I+ (I-D \Sigma)^{-1} D \Sigma\right)
\nonumber\\
&= -\Sigma \left( (I-D \Sigma)^{-1} [(I-D\Sigma) + D \Sigma]\right)
\nonumber\\\
&= -\Sigma (I-D \Sigma)^{-1},
\label{HD}
\end{align}
where the second equality follows by a matrix inversion formula: 
$(U+V)^{-1} = U^{-1} - U^{-1}(I+VU^{-1})^{-1} VU^{-1}$;
here $V$ needs not be invertible.
\qed

In what follows, we obtain explicit expressions for $(I-D \Sigma)^{-1}$
and $H^{-1}$ when the match is played on surface $m=3$.
To start, recall from \eqref{lamij} and \eqref{Lam} that
\begin{align}
\Sigma =
\begin{bmatrix}
\Lambda_1& 0\\
0 & \Lambda_2
\end{bmatrix}
=
\begin{bmatrix}
        \sigma_{11} & \sigma_{12} & \sigma_{13} & 0 & 0 & 0\\
        \sigma_{21} & \sigma_{22} & \sigma_{23} & 0 & 0 & 0\\
        \sigma_{31} & \sigma_{32} & \sigma_{33} & 0 & 0 & 0\\
        0 & 0& 0 & \tau_{11} & \tau_{12} & \tau_{13}\\
        0 & 0& 0 & \tau_{21} & \tau_{22} & \tau_{23}\\
        0 & 0& 0 & \tau_{31} & \tau_{32} & \tau_{33}
\end{bmatrix}
\label{bS}
\end{align}
and $D$ is as in \eqref{D}.
Straightforward calculation gives
\begin{align*}
I-D \Sigma &=
\begin{bmatrix}
        1 & 0& 0 & 0 & 0 & 0\\
        0 & 1& 0 & 0 & 0 & 0\\
        -d\sigma_{31} & -d\sigma_{32} & 1-d\sigma_{33} & d\tau_{31} & d\tau_{32} & d\tau_{33} \\
        0 & 0& 0 & 1 & 0 & 0\\
        0 & 0& 0 & 0 & 1 & 0\\
        d\sigma_{31} & d\sigma_{32} & d\sigma_{33} & -d\tau_{31} & -d\tau_{32} & 1-d\tau_{33} 
\end{bmatrix}
.
\end{align*}
The inverse of $I-D\Sigma$ can be obtained by Gaussian elimination.
We have 
\begin{align}
(I-D\Sigma)^{-1}=\frac{1}{{\rm det}(I-D\Sigma)}
\begin{bmatrix}
        1 & 0& 0 & 0 & 0 & 0\\
        0 & 1& 0 & 0 & 0 & 0\\
        d\sigma_{31} & d\sigma_{32} & 1-d\tau_{33} & -d\tau_{31} & -d\tau_{32} & -d\tau_{33} \\
        0 & 0& 0 & 1 & 0 & 0\\
        0 & 0& 0 & 0 & 1 & 0\\
        -d\sigma_{31} & -d\sigma_{32} & -d\sigma_{33} & d\tau_{31} & d\tau_{32} & 1-d\sigma_{33}
\end{bmatrix}
,
\label{IDSinv}
\end{align}
where ${\rm det}(I-D\Sigma)=1-d(\sigma_{33}+\tau_{33})$.
Then, plugging \eqref{bS} and \eqref{IDSinv} into \eqref{HD} gives
\begin{align*}
-H^{-1} &=\Sigma (I- D \Sigma)^{-1}\\
&= \Sigma + \frac{d}{{\rm det}(I-D\Sigma)}
\begin{bmatrix}
        \sigma_{13}\sigma_{31} &
        \sigma_{13}\sigma_{32} & \sigma_{13}\sigma_{33} &
       -\sigma_{13} \tau_{31} & -\sigma_{13}\tau_{32} &
       -\sigma_{13}\tau_{33}\\
        \sigma_{23}\sigma_{31} &  \sigma_{23}\sigma_{32}  &   \sigma_{23}\sigma_{33}  &
      - \sigma_{23} \tau_{31}  & - \sigma_{23}\tau_{32}  &
      - \sigma_{23}\tau_{33} \\
        \sigma_{33}\sigma_{31}  &  \sigma_{33}\sigma_{32}  &  \sigma_{33}\sigma_{33} & -\sigma_{33} \tau_{31}  &
        - \sigma_{33}\tau_{32}  & - \sigma_{33}\tau_{33} \\
       -\tau_{13} \sigma_{31}  & - \tau_{13}\sigma_{32}  &
       -\tau_{13}\sigma_{33}  &
         \tau_{13}\tau_{31}  &  \tau_{13}\tau_{32}  &   \tau_{13}\tau_{33} \\
      -  \tau_{23} \sigma_{31}  & - \tau_{23}\sigma_{32}  & - \tau_{23}\sigma_{33}  &
         \tau_{23}\tau_{31}  &  \tau_{23}\tau_{32}  &   \tau_{23}\tau_{33} \\
     - \tau_{33} \sigma_{31}  & - \tau_{33}\sigma_{32}  & - \tau_{33}\sigma_{33}  &
        \tau_{33}\tau_{31}  &  \tau_{33}\tau_{32} &
        \tau_{33}\tau_{33}
\end{bmatrix}
.
\end{align*}
Expressions of $-H^{-1}$ for $m=1$ and 2 can be obtained similarly.
This completes the proof of Proposition \ref{prop1}.

\section{Accuracy of \eqref{S'}}
\label{append:S'}
The Laplace method approximates the posterior distribution
of $\btheta$ with $N(\btheta^*, -H^{-1}(\btheta^*))$, where 
$\btheta^*$ is the posterior mode, $H^{-1}$ is the inverse of $H$,
and $H(\btheta^*)$
is the Hessian matrix of the logarithm of posterior distribution evaluated 
at $\btheta^*$.
Yet in \eqref{S'} the $-H^{-1}$ is evaluated at $\bmu'$,
obtained by a single Newton-Raphson step, rather than $\btheta^*$. 
In what follows we will conduct numerical studies
to assess the accuracy of the approximation in \eqref{S'}. 
Suppose that player $i$'s strength $\theta_i$
follows $N(\mu_i, \sigma_i^2)$, and that player 1 beats player 2.
We compute the approximation $(\bmu', -H^{-1}(\bmu'))$ in \eqref{S'}
and compare it with
the true posterior mean and covariance of $\btheta$ obtained by
numerical integration, denoted as $(\bmu^*, \Sigma^*)$. 
We evaluate the performance by relative errors defined as
\begin{align*}
{\rm re}(\bmu')=\frac{\|\bmu'-\bmu^*\|_2}{\|\bmu^*\|_2} \quad {\rm and} \quad
{\rm re}(-H^{-1}(\bmu'))=\frac{\|-H^{-1}(\bmu')-\Sigma^*\|_F}{\|\Sigma^*\|_F},
\end{align*}
where the $L_2$ norm of a $p$-dimensional vector $\bv=(v_1,...,v_p)^T$ 
is defined as 
$\|\bv\|_2 = (\sum_{i=1}^p v_i^2)^{1/2}$,
and the Frobenius norm of an $m \times n$ matrix A is 
$\|A\|_F = (\sum_{i=1}^m \sum_{j=1}^n a_{ij}^2)^{1/2}$.
We also compute the posterior mode $\btheta^*$ by the optimization function
{\sf optim} in {\sf R}, 
and evaluate the relative errors
of $(\btheta^*, -H^{-1}(\btheta^*))$.
Table \ref{tab:S'} reports the results using
various $(\mu_1, \mu_2)$ and $(\sigma_1, \sigma_2)$ values.
For instance, 
for $(\mu_1, \mu_2, \sigma_1, \sigma_2)=(1500, 1500, 50, 50)$,
the relative error by Laplace approximation are
$({\rm re}(\btheta^*), {\rm re}(-H^{-1}(\btheta^*)))$= 
(1.2e-06, 8.3e-04), and relative errors by a single Newton-Raphson step are
$({\rm re}(\bmu'), {\rm re}(-H^{-1}(\bmu')))$= (2.5e-06, 1.5e-02);
and with $(\mu_1, \mu_2, \sigma_1, \sigma_2)=(1500, 1800, 50, 80)$,
we have 
$({\rm re}(\btheta^*), {\rm re}(-H^{-1}(\btheta^*)))$
= (2.3e-04, 1.7e-03) and
$({\rm re}(\bmu'), {\rm re}(-H^{-1}(\bmu')))$=(2.6e-04, 5.2e-03).
In general, 
the approximate posterior means and covariances by
both $\btheta^*$ and $\bmu'$ are quite accurate,
and as expected the errors by $\btheta^*$ are lower.

\begin{table} \footnotesize
\centering
\begin{tabular}{l|c|c}
 & $(\sigma_1, \sigma_2)$=(50, 50) & $(\sigma_1, \sigma_2)$=(50, 80)\\ \hline
$(\mu_1,\mu_2)$ & Laplace \hspace{1cm} single-step & Laplace \hspace{1cm} single-step\\
(1500, 1500)& (1.2e-06, 8.3e-04) (2.5e-06, 1.5e-02)& (2.8e-05, 2.3e-03) (3.0e-05, 7.6e-03)\\
(1500, 1800)& (7.1e-05, 1.4e-03) (7.7e-05, 9.2e-03)& (2.3e-04, 1.7e-03) (2.6e-04, 5.2e-03)\\
(1500, 2000)& (4.0e-05, 1.8e-03) (4.7e-05, 4.8e-03)& (1.4e-04, 3.2e-03) (1.7e-04, 4.1e-03)
\end{tabular}
\caption{Relative errors of posterior mean and covariance approximations by
Laplace approximation $(\btheta^*, -H^{-1}(\btheta^*))$ and a single Newton-Raphson step $(\bmu', -H^{-1}(\bmu'))$}
\label{tab:S'}
\end{table}


\section*{Acknowledgements}
The authors would like to thank the Associate Editor and the referee for
their valuable comments, which helps to substantially improve the quality
of the paper.
The authors are partially supported by the National Science and Technology Council.  

\end{document}